\documentstyle[pra,aps,amsfonts,twocolumn,epsf]{revtex}  
 
\catcode`\@=11

\def\maketitle2{\par 
\begingroup
\let\cite\@bylinecite
\def\thefootnote{\fnsymbol{footnote}}%
\twocolumn[\@maketitle2\vskip2pc]%
\thispagestyle{plain}\@thanks
\endgroup
\def\thefootnote{\arabic{footnote}}%
\setcounter{footnote}{0}%
\let\maketitle2\relax \let\@maketitle2\relax
\let\@thanks\relax \let\@authoraddress\relax \let\@title\relax
\let\@date\relax \let\thanks\relax \let\@abstract\relax 
\let\@pacs\relax}

\def\abstract#1{\gdef\@abstract{{\par 
\bgroup
\ifdim\prevdepth=-1000pt \prevdepth0pt\fi
\hsize\columnwidth
\dimen0=-\prevdepth \advance\dimen0 by17.5pt \nointerlineskip
\small\vrule width 0pt height\dimen0 \relax}{~~}#1\egroup}}

\def\pacs#1{\gdef\@pacs{{\par 
\bgroup
\hsize\columnwidth \parindent0pt
\ifdim\prevdepth=-1000pt \prevdepth0pt\fi
\dimen0=-\prevdepth \advance\dimen0 by20pt\nointerlineskip
\egroup} PACS numbers:~#1}}

\def\@maketitle2{
\@preprint
\@title
\ifdim\prevdepth=-1000pt \prevdepth0pt\fi
\@authoraddress
\@date
\begin{list}{}{\leftmargin=0.10753\textwidth \rightmargin=\leftmargin
\itemsep=1pc\partopsep=-1pc}
\item\@abstract
\item\@pacs
\end{list}
}

\catcode`\@=12

\begin{document}
\draft

\title{Quantum computers in phase space}
 
\author{C\'esar Miquel$^1$, Juan Pablo Paz,$^{1,3}$ Marcos Saraceno$^2$}

\address{$^1$ Departamento de F\'\i sica, ``Juan Jos\'e 
Giambiagi'', Facultad de 
Ciencias Exactas y Naturales, UBA, Pabell\'on 1, Ciudad Universitaria, 
1428 Buenos Aires, Argentina}

\address{$^2$ Unidad de Actividad F\'{\i}sica, Tandar, CNEA  
Buenos Aires, Argentina}

\address{$^3$ Institute for Theoretical Physics, UCSB, Santa Barbara, 
CA 93106-4030, USA}

\abstract{We represent both the states and the evolution of 
a quantum computer in phase space using the discrete Wigner 
function. We study properties of the phase space representation
of quantum algorithms: apart from analyzing important 
examples, such as the Fourier Transform and Grover's search, 
we examine the conditions for the existence of a direct correspondence 
between quantum and classical evolutions in phase space. 
Finally, we describe how to directly measure the 
Wigner function in a given phase space point by means of a 
tomographic method that, itself, can be interpreted as a simple 
quantum algorithm.}

\date{\today}  
  
\maketitle2
\narrowtext

\section{Introduction}

Quantum mechanics can be formulated in phase space, the natural 
arena of classical physics. For this we can use the Wigner function 
\cite{Wigner}, which is a distribution enabling us to represent 
quantum states and temporal evolution in the classical phase space 
scenario. In this paper we use the generalization of the familiar Wigner  
representation of quantum mechanics to the case of a system with 
a finite, $N$--dimensional, Hilbert space. Our main purpose 
is to develop and study the phase space representation of both the 
states and the evolution of a quantum computer (some steps 
in this direction were described in \cite{WignerUS}). 

One can ask if there are potential advantages in using a phase space 
representation for a quantum computer. The use of this approach is 
quite widespread in various areas of physics (such as quantum 
optics, see \cite{Davidovich00} for a review) and has been fruitful,
for example, in analyzing issues concerning the classical limit of 
quantum mechanics \cite{decowigner,PazZurek00}. In answering the above
question one should have in mind that a quantum algorithm can be simply 
thought of as a quantum map acting in a Hilbert space of finite 
dimensionality (a quantum map should be simply thought of as a 
unitary operator that is applied successively to a system). 
Therefore, any algorithm is clearly amenable to a 
phase space representation. Whether this 
representation will be useful or not will depend on 
properties of the algorithm. Specifically, algorithms become 
interesting in the large $N$ limit ({\it i.e.} when operating on 
many qubits). For a quantum map this is the {\it semiclassical limit} 
where regularities may arise in connection with its classical 
behavior. Unraveling
these regularities, when they exist, becomes an important issue which 
can be naturally accomplished in a phase space representation. 
Therefore, having in mind these ideas, we conjecture that this 
representation may be useful to analyze some classes of algorithms. 
Moreover, the phase space 
approach may allow one to establish contact between the vast literature on 
quantum maps (dealing with their construction, the study of their  
semiclassical properties, etc) 
and that of quantum algorithms. This, in turn, may provide hints 
to develop new
algorithms and ideas for novel physics simulations. As a first 
application of these ideas in this paper 
we examine several properties of quantum algorithms in phase space: We
analyze under what circumstance it is possible to establish a direct
classical analogue for a quantum algorithm (exhibiting interesting
examples of this kind, such as the Fourier transform and other 
examples which naturally arise in studies of quantum maps). We also
shown that, quite surprisingly, 
Grover's search algorithm \cite{Chuang-Nielsen} can be represented 
in phase space and interpreted as a simple quantum map.

To define Wigner functions for discrete systems various attempts can be  
found in the literature. Most  
notably, Wooters \cite{Wooters} proposes a definition that has all the   
desired properties (see below) only when $N$ is a prime number. His 
phase space
is an $N\times N$ grid (if $N$ is prime) and a Cartesian product of   
spaces corresponding to prime factors of $N$ in the general case.   
Our approach here follows Wooters ideas and is closely 
related to that of Leonhardt \cite{Leonhardt} who analyzed Wigner
functions for spin systems (both in the integer and half--integer
case) and discussed interesting tomographic schemes to reconstruct the
quantum state from measurements of marginal distributions.
Other works connected to ours are those of Hannay and Berry 
\cite{Hannay}, who used discrete Wigner function in the 
context of studies of quantum chaos; Rivas and Ozorio 
\cite{Rivas} who define translation and reflection operators 
relating to the geometry of chords and centers on the
phase space torus. Bouzouina and De Bievre \cite{Bouzouina}   
give a more abstract derivation of the same Wigner function related to   
geometric quantization. 

The paper is organized as follows: 
In Section II we briefly discuss the main properties of the Wigner
representation in the continuous case. In Section III we review 
the main features of the discrete Wigner function. This 
section is supposed to be self contained and to summarize known 
results (it also contains some original results and new explanations 
of old ideas). In Section IV we examine the Wigner function 
of quantum states, some of which are of interest for quantum 
computation. In Section V we review the main properties of the temporal
evolution in the Wigner representation. Here we describe some general 
results on the nature of temporal evolution in phase space 
and draw important analogies between quantum algorithms and maps. We 
explicitly analyze Grover's algorithm in phase space and also 
discuss the conditions for quantum evolution to have a 
direct correspondence with a classical map. In Section VI we 
consider the measurement of the Wigner function. We show that
this can be done by means of a simple quantum computation that bears
a remarkable similarity with a simulated scattering experiment. 
In Section VII we present our conclusions.

\section{Continuous Wigner functions}

For a particle in one dimension, the Wigner function 
\cite{Wigner} is in one to one correspondence with the 
density matrix $\hat \rho$ and is defined as
\begin{equation}
W(q,p) = \int {{d\lambda}\over{2\pi\hbar}} \mbox{e}^{i\lambda p/\hbar}\,
\langle q-\lambda/2|\hat\rho|q+\lambda/2\rangle.
\label{wigdef}
\end{equation}
This function is the closest one can get to a phase space 
distribution for a quantum mechanical system. 
It's three defining properties are: 
(P1) $W(q,p)$ is real valued, (P2) the
inner product between states $\hat\rho_1$ and $\hat\rho_2$ 
can be computed from the Wigner function as  
$\mbox{Tr}[\hat\rho_1\hat\rho_2]=2\pi\hbar\int dq dp W_1(q,p) W_2(q,p)$ 
and (P3) the integral along any line in phase space, defined by the 
equation  $a_1 q + a_2 p = a_3$, is the probability density that a 
measurement of the observable $a_1 \hat Q + q_2 \hat P$ has $a_3$ as 
its result. This last property (the fact that $W(q,p)$ yields the 
correct marginal distribution for any quadrature) is the most 
restrictive one and, as Bertrand and Bertrand showed \cite{Bertrand}, 
together with P1--P2, uniquely determines the Wigner function. 

It is convenient to write $W(q,p)$  
as the expectation value of an operator, 
known as ``phase space point operator''  \cite{Wooters} 
(or Fano operators \cite{Fano}). 
In fact, the Wigner function is 
\begin{equation}
W(q,p) = \mbox{Tr}[\hat\rho \, \hat A(q,p) \, ].
\label{wintermsofa}
\end{equation}
The operators $\hat A(q,p)$ depend parametrically on $q$, $p$ 
and can be written in terms of simetrized products of 
delta functions as:
\begin{eqnarray}
\hat A(q,p) &=& :\delta(\hat P - p)\,\delta(\hat Q - q):\nonumber \\
&=& \int {{d\lambda d\lambda'}\over{(2\pi\hbar)^2}} \, 
\mbox{e}^{-i {\lambda\over\hbar} (\hat P-p) +i {{\lambda'}\over{\hbar}} 
(\hat Q-q)}\nonumber \\
 &=& \int {{d\lambda d\lambda'}\over{(2\pi\hbar)^2}} \, 
\hat D(\lambda,\lambda')
\, \mbox{e}^{-{i\over\hbar}(\lambda' q-\lambda p)}, 
\label{defcontA}
\end{eqnarray}
where we have identified the continuous translation operator 
$\hat D(\lambda,\lambda') = \exp[ -{i \over \hbar}(\lambda \hat P - 
\lambda' \hat Q)]$. Thus, the above expression shows that the 
phase space point operator is simply the double Fourier transform 
of the phase space displacement operator (and, therefore, $W(q,p)$ is
the double Fourier transform of the expectation value 
of $D(\lambda,\lambda')$). It is even more convenient to rewrite 
these expressions as:
\begin{equation}
\label{acontDRD}
\hat A(q,p) = {1 \over {\pi \hbar}} \hat D(q,p) \hat R 
\hat D^{\dagger}(q,p),
\end{equation}
where $\hat R$ is the reflection (parity) operator that acts on position
eigenstates as $\hat R|x\rangle = |-x\rangle$. This means that the 
Wigner function is the expectation value of a displaced
reflection operator. 
 
The proof of the three defining properties of the Wigner function (P1--P3) 
can be seen to follow from simple properties of the phase space point 
operators. The fact that $W(q,p)$ is real valued (P1) is a consequence 
of the hermiticity of $\hat A(q,p)$. Property (P2) follows from the 
completeness relation of $\hat A(q,p)$. 
In fact, one can show that these operators satisfy the relation 
\begin{equation}
\mbox{Tr}[\hat A(q,p) \hat A(q',p')]={1\over{2\pi\hbar}}
\delta(q-q')\delta(p-p').
\end{equation} 
As a consequence of this, one can invert (\ref{wintermsofa}) and express
the density matrix as a linear combination of the phase space point 
operators. The Wigner functions determines the coefficients of such
expansion: $\hat\rho=2\pi\hbar \int dq dp W(q,p)\hat A(q,p)$. From this, 
the validity of the inner product rule can be easily demonstrated. 
The last property (P3) can be seen to be valid by observing that 
integrating $\hat A(q,p)$ along a line in phase space one always gets a 
projection operator. Thus, 
\begin{equation}
\int dq dp\, \delta(a_1 q+a_2 p - a_3) \hat A(q,p) = 
|a_3\rangle\langle a_3|,
\end{equation} 
where $|a_3\rangle$ is an eigenstate of the operator 
$a_1 \hat Q + a_2 \hat P$ with eigenvalue $a_3$. This identity can be
shown by first writing the delta function as the integral of an 
exponential and then performing the phase space integration. We omit the
proof here because we will present the corresponding one for the 
discrete case, which is done following very similar lines.

\section{Discrete Wigner Functions}

\subsection{Preliminarities: discrete phase space}

We will consider here a quantum system with an $N$ dimensional 
Hilbert space (the case of a quantum computer 
is a specific example we will always have in 
mind but the formalism can be applied in other cases). 
In the Hilbert space we can  
introduce a basis $B_x=\{|n\rangle,n=0,\ldots,N-1\}$ 
which we arbitrarily  interpret as our discretized position 
basis (with periodic boundary conditions: $|n+N\rangle=|n\rangle$). 
Clearly, the notion of a position observable may not have a direct
physical interpretation in a discrete system but is 
a necessary ingredient if one wants to build a phase space 
representation. In the case of a quantum computer this 
basis can be, for example, the computational basis. Given the position 
basis $B_x$, there is a natural way to introduce the conjugate momentum 
basis $B_p=\{|k\rangle,k=0,\ldots,N-1\}$ by means of the discrete Fourier 
transform. Thus, the states of $B_p$ can be obtained from those of $B_x$ 
as
\begin{equation}
|k\rangle={1\over\sqrt{N}}\sum_n \exp(i 2 \pi nk/N)|n\rangle. 
\end{equation}
Therefore, as in the continuous case, position and momentum are
related by the Fourier transform.  It is also important to recognize
that the dimensionality of the Hilbert space of the system is related
to an effective Planck constant. In fact, phase space should have a 
finite area (which we consider equal to one, in the appropriate units). 
In this area we should be able to accommodate $N$ orthogonal states. 
If each of this states occupies a phase space area which is 
equal to $2\pi\hbar$, we see that:
\begin{equation}
N=1/2\pi\hbar. 
\end{equation}
In other words, $N$ plays the role of the inverse of the Planck 
constant (and the large $N$ limit is, in some way, the semiclassical 
limit). 

Once we have position and momentum 
basis we can construct the displacement 
operators in position and momentum. Obviously, in this case it does
not make sense to talk about infinitesimal translations. However, 
for discrete
systems we can define finite translation operators 
$\hat U$ and $\hat V$,  
which respectively generate finite translations in position and 
momentum \cite{Schwinger}. 
The translation operator $\hat U$ generates cyclic shifts in the position
basis and is diagonal in momentum basis: 
\begin{equation}
\hat U^m |n\rangle = |n+m\rangle,\ \hat U^m |k\rangle = \exp(-2\pi i mk/N)
|k\rangle,
\end{equation}
(note that all additions inside kets are to be interpreted mod $N$). 
Similarly, the operator $\hat V$ is a shift in the momentum basis and 
is diagonal in position:
\begin{equation}
\hat V^m |k\rangle=|k+m\rangle,\ \hat V^m|n\rangle = \exp (i 2 \pi mn/N)
|n\rangle
\end{equation}

It is important to notice that the translation operators $\hat U$ and 
$\hat V$ have commutation relations that directly 
generalize the ones corresponding to finite position and momentum 
translations in the continuous case. 
In fact, one can directly show the following identity:
\begin{equation}
\label{fact4}
\hat V^p \hat U^q=\hat U^q \hat V^p \exp(i 2 \pi pq/N).
\end{equation}

With these tools at hand we can introduce the discrete analog
to the continuous phase space translation operator 
$\hat D(q,p) = \exp[-{i \over \hbar}(q \hat P - p \hat Q)]$. To do this
we can first rewrite the phase space displacement as 
$\hat D(q,p)=\exp(-i q \hat P/\hbar) \, \exp(i p \hat Q/\hbar) \, 
\exp(ipq/2\hbar)$ (an expression obtained by using the well known 
formula $\mbox{e}^{\hat A + \hat B}=\mbox{e}^{-[\hat A, \hat B]/2}\,
\mbox{e}^{\hat A} \mbox{e}^{\hat B}$ and 
the canonical commutation relation between $\hat Q$ and $\hat P$). 
Therefore, by identifying 
the corresponding displacement operators, the discrete analogue of 
the phase space translation operator is:
\begin{equation}
\hat T(q,p)\equiv \hat U^q \hat V^p \exp (i \pi qp/N).
\label{Tdiscretedef}
\end{equation}
These operators obey the simple composition rule 
\begin{equation}
\hat T(q_1,p_1)\hat T(q_2,p_2)=\hat T(q_1+q_2,p_1+p_2)
{\rm e}^{i\pi(p_1q_2-q_1p_2)/N},
\nonumber
\end{equation}
where the phase appearing in the left hand side has a simple geometrical
interpretation as the area of a triangle (see below). From the above
definitions it is also simple to show that the phase space translation
operators are such that
\begin{eqnarray}
\hat T^\dagger(q,p)&=& \hat T(2N-q,2N-p)\label{tdaggerp2}\\
 &=& \hat T(N-q,N-p) \, (-1)^{N+p+q} \label{tdaggerp1}. 
\end{eqnarray}
Finally, it is worth mentioning the fact that the composition rule can 
be used to show that 
\begin{equation}
\hat T(\lambda q, \lambda p) = \hat T^\lambda(q,p) 
\label{tlambda}
\end{equation}
for any integer $\lambda$, which is a very natural result. 
It may be tempting to define position and momentum operators also in 
the discrete case. In fact, we could attempt doing that by writing 
$\hat U$ and $\hat V$ as the exponentials of
two operators $\hat Q$ and $\hat P$ defined to be diagonal in $B_x$ 
and $B_p$ respectively: $\hat U=\exp(-i 2 \pi \hat P/N)$ 
and $\hat V=\exp(i 2 \pi \hat Q/N)$, with 
$\hat Q \equiv \sum_n n |n\rangle\langle n|$ and 
$\hat P \equiv \sum_k k |k\rangle\langle k|$. However, it
turns out that these operators are not canonically conjugate since
their commutator {\it is not} proportional to the identity as is the 
continuous case (actually, it is well known that for finite dimensional
spaces it is impossible to find two operators whose commutator is 
proportional to the identity). 
So, even though finite shifts in the discrete case have the same 
properties than their continuous counterparts, this is not true
for the position and momentum operators. 

It will also be useful to define a reflection (or parity) operator 
$\hat R$ as the one acting on states in the position (and momentum) 
basis as $\hat R|n\rangle \equiv |-n\rangle$ 
(again, the operation is to be understood mod $N$). It is simple to 
show that the parity operator obeys simple relations with the shifts
$\hat U$ and $\hat V$: 
\begin{equation}
\label{fact3}
\hat U \hat R = \hat R \hat U^{-1}, \ \hat V \hat R = \hat R \hat V^{-1}. 
\end{equation}
It is important to recognize that the reflection operator is directly
related to the Fourier transform. In fact, if we denote the operator
implementing the discrete Fourier transform as $U_{FT}$ (i.e., the 
operator whose matrix elements in the $B_x$ basis are 
$\langle n'| U_{FT}| n\rangle = \exp(i2\pi n n'/N)$), then the reflection
operator is simply the square of the Fourier transform:
\begin{equation}
\label{R=UFT^2}
\hat R=U_{FT}^2.
\end{equation}

Before ending this section it is worth mentioning 
other possibilities for the phase space description of finite 
quantum mechanics. The use of the discrete Fourier transform to 
relate position and momentum implies the imposition of periodic 
boundary conditions on both these variables, thus imposing the 
geometry of a torus on the phase space. This is by no means 
mandatory, as the well known example of the angular momentum 
sphere --spanned by states $|j,m\rangle , m=-j,...j$-- 
illustrates. However, the ubiquitous appearance of the DFT 
in quantum algorithms somehow favors the choice of the torus 
as the preferred geometry. It is also possible to slightly extend 
the scope of the present approach by imposing {\it quasi-periodic} 
boundary conditions, as is done in solid state chain models or 
in the treatment of the quantum Hall effect. This generalization 
can be easily incorporated into the description of quantum 
algorithms \cite{CircuitsUS}

\subsection{Discrete Wigner functions and phase space point operators}

To define a discrete Wigner function the most convenient strategy is
to find the correct generalization of the phase space point operators
$\hat A(q,p)$. Thus, we need to find a basis set of the space
of hermitian operators 
(note that, as the Hilbert space is $N$ dimensional such a basis has
$N^2$ independent operators). We should do this in such a way that 
the resulting Wigner 
function gives the correct marginal distributions.
However, generalization from the continuous to the discrete case 
has to be done with some caution. In fact,
it is instructive to see how naive attempts to generalize expressions like
(\ref{defcontA}) or (\ref{acontDRD}) fail. In fact, generalizing 
equation (\ref{defcontA}) to the discrete case would lead us to define
\begin{equation}
\label{AgridN}
\hat{\cal A}(q,p)={1\over N^2}\sum_{m,k=0}^{N-1}
\hat T(m,k)\exp{\{-i2\pi{(kq-mp)\over N}\}}.
\end{equation}
Unfortunately, this expression for $\hat {\cal A}(q,p)$ turns out to be
non--hermitian. This can be seen by taking the hermitian conjugate of 
the above equation and using the fact that the Hermitian conjugate of 
the phase space translation operator, as shown in (\ref{tdaggerp1}), 
is such that $\hat T^\dagger(q,p)\neq \hat T(N-q,N-p)$. Therefore, the 
above is not a good definition for the discrete phase space point operators. 

Another naive attempt to define discrete phase space operators is to 
generalize equation (\ref{acontDRD}) which tells us that in the 
continuous case the operator $\hat A(q,p)$ is a displaced reflection.  
Thus, by writing
\begin{equation}
\label{definition_of_a1}
\hat A(q,p) = {1 \over {\pi \hbar}} \hat D(q,p) \hat R \hat D^{\dagger}(q,p) 
\rightarrow {1\over N/2} \hat T(q,p) \hat R \hat T^{\dagger}(q,p), 
\end{equation}
one defines an operator that is hermitian by construction. However, in this
case the problem is that the number of operators one defines in this
way is not enough to build a complete set. In fact,  
if one uses the definition of $\hat T(q,p)$ given in (\ref{Tdiscretedef})
and then the commutation relations between $\hat U$, $\hat V$ and 
$\hat R$ one finds:
\begin{equation}
\label{Anogo2}
\hat A(q,p) = {1\over N} \,\hat U^{2q} \hat R \hat V^{-2p} 
\exp(4\pi i \,pq/N).
\end{equation}
As $\hat U$ and $\hat V$ are cyclic operators with period $N$, 
it is easy to see that if both $q$ and $p$ take values between $0$ 
and $N$ the above expression only gives $N^2/4$ independent operators.  
In fact, it is clear that the above equation implies that
$\hat A(q+N/2,p)=\hat A(q,p)$ and likewise with $p$ so we only have 
$N/2 \times N/2$ independent operators. 

Remarkably, the solution to the problems arising with the two above 
naive definitions is obviously the same: We should define, as it is 
done in the literature \cite{Hannay,Bouzouina} the phase
space point operators on a phase space grid of $2N\times 2N$ points. 
This can be done simply replacing $N\rightarrow 2N$ in 
equation (\ref{AgridN}) or by taking half integer values for $q$ and $p$
in (\ref{Anogo2}). The two definitions turn out to be equivalent. 
Therefore, the correct discrete phase space point operators are 
\begin{eqnarray}
\label{definition_of_a2}
\hat A(\alpha)&=&{1\over (2N)^2}\sum_{\lambda,\lambda'=0}^{2N-1} 
\hat T(\lambda,\lambda')\exp{\{-i2\pi{(\lambda' q -\lambda p)\over 2N}\}}
\label{disphase1}\\ 
&=& {1 \over {2N}} \hat U^q \hat R \hat V^{-p} \exp(i \pi pq/N)
\label{disphase2}
\end{eqnarray}
where $\alpha=(q,p)$ denotes a point in the phase space grid with
$q$ and $p$ taking values between $0$ and $2N-1$. 
 
It is worth noticing that, as we have just defined phase space 
point operators on a lattice with $2N \times 2N$ points, we have 
a total of $4N^2$ such operators. However, it should be clear that 
these operators are not all independent. It is easy to verify that 
there are only $N^2$ independent phase space point operators 
since it can be proved that:
\begin{equation}
\label{Arelations}
\hat A(q+\sigma_q N, p+\sigma_p N)=
\hat A(q,p) \, (-1)^{\sigma_p q + \sigma_q p + \sigma_q \sigma_p N}, 
\end{equation}
for $\sigma_q,\sigma_p=0,1$. 
Therefore, it is clear that the $N^2$ phase space point operators 
corresponding to the first $N\times N$ subgrid of the phase space 
determine the rest. For the rest of the paper we will denote the 
first $N\times N$ subgrid as the set $G_N$ (i.e., $G_N=\{\alpha=(q,p);
0\le q,p\le N-1\}$). The set $G_{2N}$ will denote the full $2N\times 2N$ 
grid. 

Before showing explicitly that the above operators enable us to define
a Wigner function with all the desired properties, it is worth pointing
out some useful facts about $\hat A(\alpha)$. In particular, it is 
convenient to realize that these operators are closely related to 
the phase space translation operators. In fact, one can show that 
by successive application of two phase space operators one always gets
a translation (this is also true in the continuous case) since 
\begin{equation}
\label{Acomposition}
\hat A(\alpha) \hat A(\alpha') = 
 \hat T(\alpha -\alpha') \, {{\mbox e}^{i{\pi\over N}
(q_\alpha p_{\alpha'}-q_{\alpha'}p_\alpha)}\over{4N^2}}
\end{equation}
Moreover, it is also useful to express the translation
operator in terms of the $\hat A(\alpha)$ operators by simply 
inverting the defition (\ref{disphase1}) and writing 
\begin{equation}
\tilde T(n,k)=\sum_{q,p=0}^{2N-1} \, \hat A(q,p) \ 
\exp(-i {2\pi\over{2N}}(np-kq)),
\label{ftofa}
\end{equation}
a property that is also valid in the continuous case. 

From the above properties it is possible to show that $\hat A(\alpha)$ 
are a complete set when $\alpha$ takes values in the first $N\times N$
subgrid $G_N$. Thus, taking the trace of (\ref{Acomposition}) one
gets:
\begin{equation}
\mbox{Tr}[\hat A(\alpha) \hat A(\alpha') ] = 
{1\over {4N}} \, \delta_N(q'-q) \delta_N(p'-p) 
\end{equation}
where both $\alpha$ and $\alpha'$ are in the grid $G_N$ and 
$\delta_N(q) \equiv {1\over N}\sum_{n=0}^{N-1} \mbox{e}^{2 \pi i q/N}$ 
is the periodic delta function (which is 
zero unless $q=0 \ \mbox{mod} \ N$). 

Therefore, according to all the previous arguments, the discrete 
Wigner function is defined as
\begin{equation}
\label{wigdefdisc}
W(\alpha)=\mbox{Tr}(\hat A(\alpha)\hat \rho)
\end{equation}
where $\alpha\in G_{2N}$. Clearly, these
$4N^2$ values are not all independent since the Wigner function obeys
the same relation than the phase space point operators:
\begin{equation}
\label{Wrelations}
\hat W(q+\sigma_q N, p+\sigma_p N)=
\hat W(q,p) \, (-1)^{\sigma_p q + \sigma_q p + \sigma_q \sigma_p N} 
\end{equation}

As the operators $\hat A(\alpha)$ are a complete set 
one can expand the density matrix
as a linear combination of such operators. It is clear that the 
Wigner function $W(\alpha)$ are nothing but the coefficients of such
expansion. Thus, one can show that 
\begin{eqnarray}
\hat \rho &=& 4N \,  \sum_{\alpha\in G_N} W(\alpha) \hat A(\alpha) 
\label{rhofromwN}\\
&=& N \,  \sum_{\tilde\alpha\in G_{2N}} W(\tilde\alpha) \hat 
A(\tilde\alpha). 
\label{rhofromw2N}
\end{eqnarray}
The last expression can be obtained from (\ref{rhofromwN})
by noticing that the contribution arising from the four $N\times N$
subgrids are identical. 

Now it is simple to show that the Wigner function defined above 
obeys the three defining properties (P1--P3). The first one is
a consequence of the fact that the operators $\hat A(\alpha)$ are
hermitian by construction. The second property (P2) is a consequence
of the completeness of the set $\hat A(\alpha)$ which enables one to 
show that 
\begin{equation}
\mbox{Tr}[\rho_1 \rho_2]=N \sum_{\alpha\in G_{2N}} 
W_1(\alpha) W_2(\alpha)
\end{equation}

\begin{figure}
 \centering \leavevmode
 \epsfxsize 3.3in
 \epsfbox{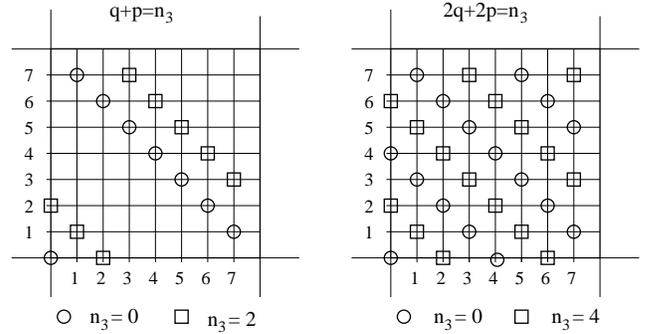}
\caption{Left: lines satisfying the equation $p+q=n_3$ (mod $8$) 
for values $n_3= 0, 2$. Each line has 8 points and there are a 
total of 8 distinct lines. Right: two of the four possible lines 
satisfying $2p+2q=n_3$ with $n_3=0, 2$. In both plots horizontal
(vertical) axis corresponds to position (momentum) basis. 
}
\label{linesongrid}
\end{figure}

Finally, the third property (P3) is more subtle and deserves to be
studied with some detail. The crucial point is to show that if we
add the operators $\hat A(q,p)$ over all phase space points lying on 
a line $L$ we always obtain a projection operator (this guarantees 
that adding the value of the Wigner function over all points in a line
gives always a positive number, which can be interpreted as a 
probability). Before showing this, we should clearly
define what do we mean by a line $L(n_1,n_2,n_3)$ 
on our phase space grid. We will 
use the following definitions: A line $L$ is a set of points of the 
grid defined as $L=L(n_1,n_2,n_3)=\{(q,p)\in G_{2N},\ \mbox{such that} 
\ n_1 p - n_2 q = n_3$, \mbox{with}\ $0\le n_i\le 2N-1\}$. Below, 
we will discuss some more details about the structure of 
lines on a grid $G_{2N}$. Here, we only need to point out that one
can also define a notion of parallelism between lines: two lines 
parametrized by the same set of integers $n_1$ and $n_2$ are parallel
to each other (in Figure (\ref{linesongrid}) we show 
some examples of lines on a grid). 

So, let us show that by adding phase space point operators on a line 
one always gets projection operators. We are interested in looking
at the operator $\hat A_L$ defined as 
\begin{equation}\label{ALdef}
\hat A_L = \sum_{(q,p)\in L} \hat A(q,p).
\end{equation}
It is clear that this operator can be rewritten as
\begin{eqnarray}
A_L &=& \sum_{q,p=0}^{2N-1} \ \hat A(q,p) \ 
\delta_{2N}(n_1 p - n_2 q - n_3) \nonumber\\
 &=& {1\over 2N}
\sum_{\lambda=0}^{2N-1} \ \sum_{q,p=0}^{2N-1} \ \hat A(q,p) \ 
\mbox{e}^{-i {2\pi \over {2N}} \lambda (n_1p - n_2 q - n_3)}\nonumber \\
 &=& {1\over 2N} \ \sum_{\lambda=0}^{2N-1} \hat T^\lambda(n_1,n_2)
\ \mbox{e}^{i {2\pi\over{2N}} n_3\lambda} \label{AL2}
\end{eqnarray}
where to perform the sum over the phase space grid we used 
the fact that the Fourier transform of $\hat A(\alpha)$ is given
in (\ref{ftofa}). Now, as  
$\hat T(n_1,n_2)$ is unitary, it has $N$ eigenvectors
$|\phi_j\rangle$ with eigenvalues $\exp(-i 2\pi\phi_j/N)$. Moreover, 
this operator is cyclic and satisfies
$\hat T^N={\Bbb I}$. Therefore, as its eigenvalues are $N$th roots of 
unity, $\phi_j$ must be integers. Then, we can rewrite 
equation (\ref{AL2}) as:
\begin{eqnarray}
\hat A_L &=& {1\over 2N} \sum_{\lambda=0}^{2N-1} \ \sum_{j=0}^{N} 
\mbox{e}^{-i {{2\pi}\over{2N}}(2\phi_j-n_3)\lambda} \ 
|\phi_j\rangle\langle\phi_j| \nonumber\\
&=&  \ \sum_{j=0}^{N} \delta_{2N}(2\phi_j-n_3) \ 
|\phi_j\rangle\langle\phi_j|.  
\end{eqnarray}
Therefore, it is clear that $\hat A_L$ is a projection operator 
onto a subspace generated by a subset of the 
eigenvectors of the translation operator $\hat T(n_1,n_2)$.
The dimensionality $d$ of this subspace is equal to the trace 
of $\hat A_L$. To calculate it, we just have to notice that in 
general
\begin{equation}
\mbox{Tr}[\hat A(q,p)] =
{1\over {2 N}} \sum_{q'=0}^{N-1} \delta_N(q-2q') \ 
\mbox{e}^{i \pi (q-2q')p/N}
\end{equation}
This last equation is easy to evaluate but the result depends on 
the parity of $N$. 
Thus, if $N$ is even then $\mbox{Tr}(\hat A(q,p))=1/N$
if both $q$ and $p$ are even and it is zero otherwise. On the other hand
for odd values of $N$, one has that $\mbox{Tr}(\hat A(q,p))=1/2N$ for 
all values of $q$ and $p$ except when they are both odd where it has the
opposite sign (i.e., it is equal to $-1/2N$). 
In the rest of the paper we will concentrate on analyzing the case of 
even $N$ which is somewhat more relevant for a quantum computer (that
has at least one qubit and therefore a Hilbert space which is even 
dimensional). Using the above result for the trace of 
$\hat A(\alpha)$ one can see that the dimensionality of the 
projector $\hat A_L$ is simply given by
$1/N$ times the number of points that belong to the line $L$ which 
have even $q$ and even $p$ coordinates. An immediate consequence
of this is that if $n_3$ is odd then $d=0$ (since the sum of two even
numbers can never be odd). Finally we can write an explicit formula
for $d$ in the case of even $N$:
\begin{equation}
d = {1\over 2} \, \sum_{\lambda=0}^{N-1} \ \delta_N(\lambda n_1) 
\delta_N(\lambda n_2) \mbox{e}^{i {2\pi m\lambda\over 2N}} 
\, (1+(-1)^{n_3})
\end{equation}

Let us illustrate the above result applying it 
to the simplest example: For a line $L_q$ defined  
as $q=n_3$ (i.e., $n_1=1$, $n_2=0$), the Wigner function summed over   
all points in $L_q$ is $\sum_{(q,p)\in L_q} W(q,p)=\sum_p W(n_3,p)=  
\langle n_3/2|\hat\rho|n_3/2\rangle$ if $n_3$ is even  
(and zero otherwise). Analogously, considering horizontal lines
($L_p$ defined as $p=n'_3$) we get  
$\sum_{(q,p)\in L_p} W(q,p)=\sum_q W(q,n'_3)=  
\langle n'_3/2|\hat\rho|n'_3/2\rangle$ if $n'_3$ is even  
(and zero otherwise). These are just two examples a general result: 
this Wigner function always generates the correct marginal 
distributions (this, as in the 
continuous case, is the defining feature of $W(q,p)$).  


As a final point in this section where we 
reviewed some known (and other not so well known) 
results on discrete Wigner functions we think it is convenient to 
mention some simple properties of the phase space grid $G_{2N}$. 
Most of the ideas we are using were introduced by Wootters \cite{Wooters}.
As mentioned above we can simply define lines $L(n_1,n_2,n_3)$ and 
introduce a notion of parallelism in the grid $G_{2N}$. A foliation
of the grid with a family of parallel lines is obtained by fixing
$n_1$ and $n_2$ and varying $n_3$ (in general, it is evident that 
two lines are parallel if the ratio $n_1/n_2$ is the same). 
If $N$ is a prime number then in the grid $G_N$ (which has $N\times N$
points) there are exactly $N(N+1)$ distinct lines which can be grouped 
into $N+1$ sets of parallel lines (these are $N+1$ different 
foliations of the grid). If $N$ is not prime or, as it is the case we
are interested in, in the grid $G_{2N}$ 
this result is no longer true. For example, it is very clear that 
the equation $n_1 q-n_2 p=n_3$ (mod $2N$) has no solutions for odd 
values of $n_3$ and even values of $n_1$ and $n_2$. So, it is not
generally true that our lines have always $2N$ points: As shown above
sometimes they can have no points at all and in some cases one can 
construct lines with a number of points that is a multiple of $2N$. 
This is the case if $n_1$ and $n_2$ have common prime 
factors with $N$. For example in the case $2N=8$, each line with 
$n_1=n_2=1$ have $8$ points, but lines with $n_1=n_2=2$ have no points
if $n_3$ is odd or $16$ points each when $n_3$ is even. 
In any case, the simplest way to represent the lines in the phase 
space is by noticing that, as the topology of the lattice is that
of a torus (due to the cyclic boundary conditions) lines wrap around
the torus. The number of points in the line is related to the number
of times it wraps around before it closes onto itself. 
Figure \ref{linesongrid} shows an example of two sets of lines on 
the grid $G_{2N}$ for the case of $N=4$. 

Let us summarize the results presented in this Section: We 
defined a Wigner function for systems with a finite dimensional 
Hilbert space (of arbitrary dimension $N$). The Wigner function  
is defined as the expectation value of the phase space operator
$\hat A(\alpha)$ given in (\ref{disphase2}). This definition is 
such that $W(\alpha)$ is real, it can be used to compute
inner products between states and it gives all the correct 
marginal distributions when added over any line in the phase 
space, which is a grid $G_{2N}$ with $2N\times 2N$ 
points. The size of the phase space grid is important to obtain 
a Wigner function with all the desired properties. The values of  
$W(\alpha)$  on the subgrid $G_N$ are enough to reconstruct 
the rest of the phase space (since the set $\hat A(\alpha)$ is 
complete when $\alpha$ belong to the grid $G_N$). However, the 
redundancy introduced by the doubling of the number of sites in 
$q$ and $p$ is essential when one imposes the condition that 
all the marginal distributions should be obtained from the Wigner
function. In the rest of the paper we will concentrate on 
applying this Wigner function to study the states and the 
evolution of a typical quantum computer.

\section{Wigner Functions of quantum states}

To compute the Wigner function of any quantum state it is 
convenient to use equation (\ref{disphase2}) for $\hat A(q,p)$
and write:
\begin{equation}\label{Wsimple}
W(q,p) \equiv {1 \over {2{N}}} 
\sum_{n=0}^{N-1} \langle q-n|\hat\rho|n\rangle 
\mbox{e}^{i {2 \pi\over N} n(p-q/2)}.
\end{equation}
Moreover, it is worth remembering that one only needs to compute
$N^2$ independent values (restricting to $\alpha=(q,p)\in G_N$) 
and from them one can reconstruct the remaining $3N^2$ 
ones by using (\ref{Wrelations}).
Before showing the specific form of the Wigner function of some 
states it is worth mentioning some general 
features of the Wigner function of pure states. In fact, if the 
quantum state is pure then $\hat\rho$ is a projection operator. Then, 
expanding $\hat\rho$ in terms of the phase space operators as in
(\ref{rhofromw2N}) and imposing the condition $\hat\rho^2=\hat\rho$
one gets 
\begin{equation}
\label{Wpure}
W(\alpha)=4N^2\sum_{\beta\gamma\in G_N}\Gamma(\alpha,\beta,\gamma)
W(\beta)W(\gamma).
\end{equation}
where the function $\Gamma(\alpha,\beta,\gamma)$, that depends on three
phase space points (i.e, on a triangle) is
\begin{eqnarray}
\Gamma(\alpha,\beta,\gamma)&=& \mbox{Tr}(\hat A(\alpha)\hat A(\beta)
\hat A(\gamma))\nonumber\\
&=& {1\over 4N^3} \mbox{e}^{i {2\pi\over N}  
{\sl S}(\alpha,\beta\gamma)}
\label{Gamma}
\end{eqnarray}
if either two or three of the points $(\alpha,\beta,\gamma)$ have
even $q$ and even $p$ coordinates. Otherwise 
$\Gamma(\alpha,\beta,\gamma)=0$. In the above expression, which
is valid for even values of $N$, ${\sl S}(\alpha,\beta,\gamma)$ is the
area of the triangle formed by the three phase space points (measured
in units of the elementary triangle formed when the three points are 
one site apart). A similar expression is obtained by Wooters \cite{Wooters}. 
The three point function $\Gamma(\alpha,\beta,\gamma)$ has then a 
simple geometric meaning and will play an interesting role in determining
the properties of the temporal evolution (as Unitary evolution preserves
pure states, it will be represented by a linear map in phase space that 
leaves $\Gamma$ invariant, as shown see below).

\subsection{Position and momentum eigenstates, and their superpositions}

Here, we first evaluate the Wigner function of a position 
eigenstate (a computational state of the quantum computer) 
$\hat\rho_{q_0} = |q_0\rangle\langle q_0|$. It is straightforward 
to obtain a closed expression for $W(q,p)$:
\begin{eqnarray}
W_{q_0}(q,p) 
 &=& {1\over {2 N}} \langle q_0|\hat U^q \hat R 
\hat V^{-p} |q_0 \rangle \mbox{e}^{i\pi  pq/N} \nonumber\\
 &=& {1 \over {2N}} \delta_N(q-2 q_0) \ (-1)^{p(q-2q_0)_N}
\end{eqnarray}
where $z_N$ denotes $z$ modulo $N$. This function is zero only on
two vertical strips located at $q=2q_0$ modulo $N$. When $q=2q_0$, 
$W(q,p)$ 
takes the constant value $1/2N$ while for $q=2q_0 \pm N$ it is $1/2N$ for
even values of $p$ and $-1/2N$ for odd values. These oscillations are
typical of interference fringes and can be interpreted as arising 
from interference between the $q=2q_0$ strip and a mirror image formed 
at a distance $2N$ from $2q_0$, which is induced by the 
periodic boundary conditions. The
fact the the Wigner function becomes negative in this interference strip
is essential to recover the correct marginal distributions. Adding
the values of $W(q,p)$ along a vertical line gives the probability of
measuring $q/2$ which should be $1$ for $q=2q_0$ and zero otherwise.
A very similar calculation can be done for a momentum eigenstate 
$\hat\rho = |k_0\rangle \langle k_0|$. The result is very similar except
that now the strips are horizontal. 

\begin{figure}
 \centering \leavevmode
 \epsfxsize 3.2in
 \epsfbox{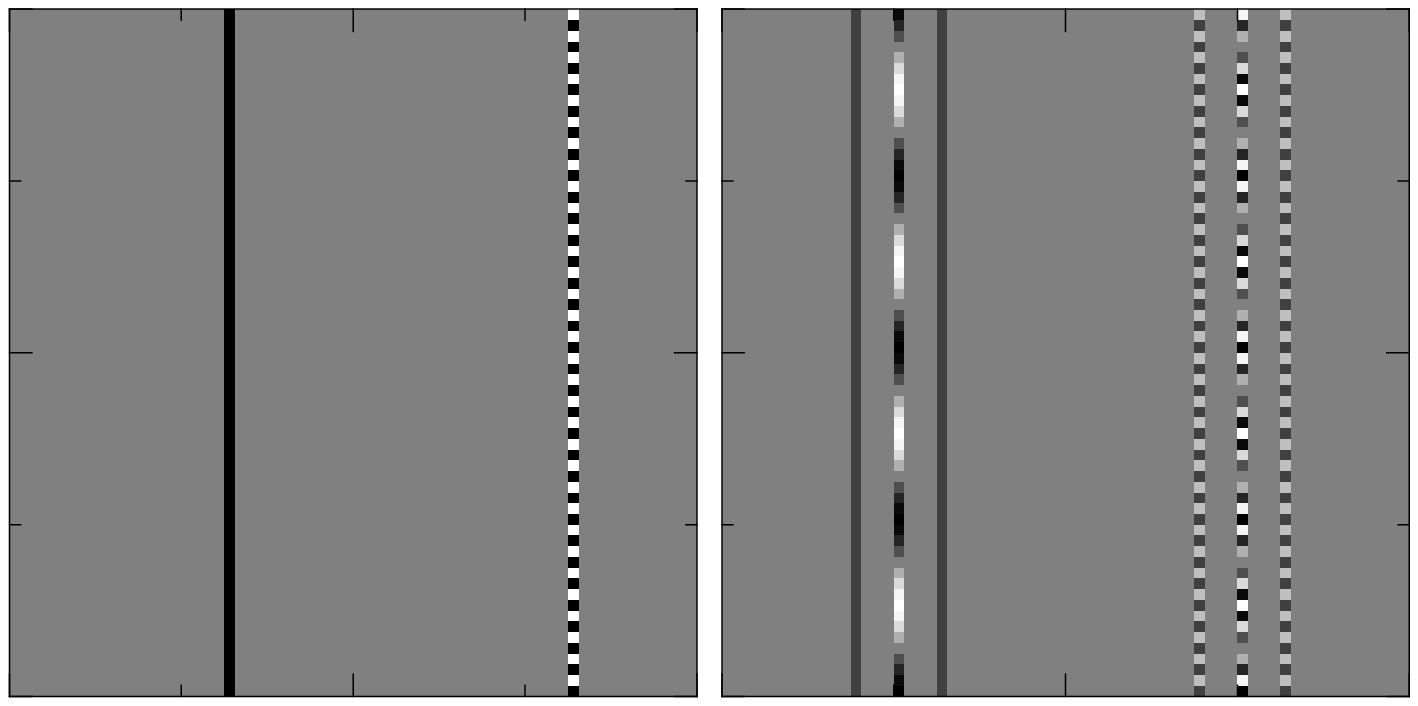}
\vspace{0.25 cm}
\caption{Wigner function of a position eigenstate (left) 
and a linear superposition of two position 
eigenstates (right). Horizontal (vertical) axis correspond
to position (momentum) basis. For the computational state $|q_0\rangle$, 
the Wigner function is positive (black) on a vertical strip 
located at $q=2q_0$ and it has negative (white) values on 
another vertical strip located at $q=2q_0\pm N$. For a superposition
of two computational states (right), the Wigner function also 
oscillates on the strip located at $q=q_0+q_1$ 
with a wavelength that depends upon the distance between the two 
interfering states.}
\label{Wignerqsuper}
\end{figure}

It is interesting to analyze also the Wigner function of a state which 
is a linear superposition 
$|\psi\rangle = \left( |q_0\rangle + \mbox{e}^{-i \phi}
|q_1\rangle\right)/\sqrt{2}$. Again one can simply get a closed 
expression for $W(q,p)$ which is 
$$
W(q,p)={1\over 2} \left[ W_{q_0}(q,p) + W_{q_1}(q,p) + 
\Delta W_{q_0,q_1}(q,p) \right]\nonumber
$$
where the interference term is
$$
\Delta W_{q_0,q_1}(q,p) \equiv {1 \over N} \,
\delta_N(\tilde q) \, (-1)^{\tilde q p} \, \cos \left( {{2\pi} 
\over \lambda}p + \phi \right), \nonumber
$$
with $\tilde q = q_0+q_1-q$ 
and $\lambda \equiv 2N/(q_0-q_1)$. Thus, this Wigner function has
two direct terms which simply correspond to the two computational
states and an interference term which is peaked on 
the vertical strips located at the midpoint $q=q_0+q_1 \ \mbox{mod} \ N$. 
On this strip the Wigner function oscillates with a wavelength that 
is inversely proportional to the separation between the main fringes. 
This oscillatory pattern has its corresponding mirror image and the 
interference between them produce some more oscillations. 
In figure \ref{Wignerqsuper} we show the Wigner function for a 
state and a superposition of two such computational 
states. The plot is much more eloquent than any equation and
shows both the presence of the main peaks, its mirror images and (in 
the case of a superposition state) the interference fringes. 

\subsection{Other quantum states}

\begin{figure}
 \centering \leavevmode
 \epsfxsize 3.2in
\epsfbox{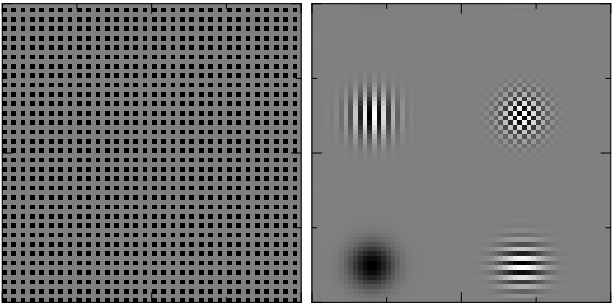}
\vspace{0.25 cm}
\caption{Left: Wigner function for the completely mixed state. It
is positive (black) only in the phase space points with even values
of $q$ and $p$ being zero (grey) elsewhere. Right: Wigner function for
a periodic Gaussian wavepacket. This function has a positive (black)
peak and three regions where it displays oscillatory behavior. These
regions can be understood as being originated from the interference
between the positive Gaussian peak and its mirror images created
by the periodic boundary conditions. Horizontal (vertical) axis correspond
to position (momentum) basis.}
\label{cohunit}
\end{figure}

The Wigner function can be computed in closed form for some other
interesting states. In figure \ref{cohunit} we display such 
function for the completely mixed state (that, as we mentioned above, 
is nonzero only when both $q$ and $p$ are even). We also show the 
Wigner function for a pure state constructed as a Gaussian 
superposition of computational states (with periodic boundary 
conditions, i.e. a sum of a Gaussian state centered about a 
phase space point and all its infinite mirror images). The Gaussian 
wavepacket has a Wigner function with a positive Gaussian peak and
shows three other Gaussian peaks that are modulated by 
interference fringes. They correspond to the interference between the 
main peak and its mirror images (notice the different orientation of 
these fringes).

\section{Temporal evolution in phase space}
 
Temporal evolution of a quantum system can also be represented
in phase space. If $U$ is the unitary evolution operator
that takes the state of the system from time $t$ to time $t+1$, 
the density matrix evolves as
\begin{equation}
\rho(t+1)=U\rho(t) U^\dagger.\nonumber
\end{equation}
Using this, it is simple to show that 
the Wigner function evolves as
\begin{equation}
W(\alpha,t+1)=\sum_{\beta\in G_{2N}} Z_{\alpha\beta} W(\beta,t).
\label{wignert+1}
\end{equation}
where the matrix $Z_{\alpha\beta}$ is defined as 
\begin{equation}
Z_{\alpha\beta}=N\quad Tr(\hat A(\alpha) U \hat A(\beta) U^\dagger).
\label{zeta}
\end{equation}
Therefore, temporal evolution in phase space is represented by 
a simple linear transformation (which is, of course, an immediate 
consequence of linearity of Schr\"odinger equation). Unitarity 
imposes some constraints on the matrix $Z_{\alpha\beta}$: As 
purity of states is preserved, temporal evolution must 
preserve the structure of the constraint equation (\ref{Wpure}). 
Therefore, the matrix must leave invariant the three point 
function $\Gamma(\alpha,\beta,\gamma)$, i.e.
\begin{equation}
\Gamma(\alpha',\beta',\gamma')=\sum_{\alpha\beta\gamma}
Z_{\alpha'\alpha}Z_{\beta'\beta}
Z_{\gamma'\gamma}\Gamma(\alpha,\beta,\gamma).\nonumber
\end{equation}

The real matrix $Z_{\alpha\beta}$ contains all the information about
the nature of temporal evolution. In general, this matrix 
connects a point $\alpha$ with many other points $\beta$. Therefore, temporal 
evolution will be generally nonlocal in phase space. This is a unique 
quantum mechanical feature: In fact, classical systems evolve in phase 
space following a flow of classical trajectories. In such case, the value 
of the classical distribution function $W(\alpha,t+1)$ is 
equal to the value $W(\beta,t)$ for some point $\beta$ which is a well 
defined function of $\alpha$ and $t$. One may ask what kind of 
Unitary operators generate a local dynamical evolution in phase
space. Below, we will give examples of this type. 

\subsection{Phase space translations}

It is simple to show that if we consider the unitary evolution 
to be the phase space translation operator 
$\hat T(\sigma)=\hat T(q,p)$ then temporal 
evolution is local in phase space. Moreover, quantum and classical 
evolution are identical in this case since the value of the 
Wigner function is rigidly translated in phase space: 
\begin{equation}
U=\hat T (\sigma)\iff W(\alpha,t+1)=W(\alpha-2\sigma,t). 
\label{wtranslation}
\end{equation} 
Notice that the factor of $2$ in the above equation is simply a 
consequence of the fact that we are working in a phase space grid
of $2N\times 2N$. Another example of a local evolution is the one 
associated with the phase space operators $\hat A(\alpha)$ 
themselves. In such case the resulting evolution is not just 
a translation but a translation combined with a reflection:
\begin{equation}
U=\hat A (\sigma)\iff W(\alpha,t+1)=W(2\sigma -\alpha,t). 
\label{atranslation}
\end{equation} 
The above are two simple examples of families of unitary 
operators with corresponding local phase space representations. 
As we will see below, the family of operators with a direct 
correspondence between quantum and classical evolution has 
other interesting members. 

\subsection{Fourier transform}

The discrete Fourier transform is a unitary operator that is 
widely used in quantum algorithms \cite{Chuang-Nielsen}. 
As position and momentum 
basis are related by this operator one expects that it should
have a rather simple phase space representation. In fact, this
is indeed the case: One can easily show that the Fourier transform
is represented as a rotation in $90$ degrees:
\begin{equation}
U=U_{FT} \iff W(q,p,t+1)=W(-p,q,t). 
\label{wft}
\end{equation} 
where $-p$ is the additive inverse mod$(N)$. 
Thus, the Fourier transform is also represented by a local 
operation in phase space (and, in this sense it is a completely
``classical'' operation, acting independently on each of the $N\times
N$ subgrid). For example, by applying the Fourier
transform to the quantum states whose Wigner functions are shown
in Figure 2 (a computational state and a superposition of two such
states) one obtains the resulting Wigner function by rotating 
the above figure by $90$ degrees (i.e., one gets a momentum eigenstate
or a superposition of two such states where the vertical pattern
in Figure 2 is mapped into the same horizontal pattern). It is clear
that by applying the Fourier transform twice one gets a rotation by 
$180$ degrees. This is nothing but a reflection, which is what 
equation (\ref{R=UFT^2}) tells us. 

\subsection{Quadratic (cat) maps propagate classically}

The above operators (rigid translations, reflections and the 
Fourier transform) are just some examples of unitary operators 
which generate a classical evolution in phase space. A more general 
family with such property will be discussed here. This family
is formed by the quantization of classical dynamical systems with 
a quadratic Hamiltonian and a finite phase space with periodic
boundary conditions (as the phase space is a torus, the classical 
transformations are the linear automorphisms of the torus \cite{Arnold}). 
We will briefly present these unitary operators here and 
show that classical and quantum evolution coincides (a fact 
that has been shown before, using different techniques, 
by Hannay and Berry \cite{Hannay}). 
Consider the following two-parameter family of operators
(it is possible to consider a slightly more general case, with
one more parameter, but to simplify the presentation we 
restrict to this simpler case \cite{CircuitsUS}).
\begin{equation}
U_{cat}={\cal V}_b{\cal T}{\cal V}_a.\label{operatorcat}
\end{equation}
The operators ${\cal V}_a$ and ${\cal T}$ are 
respectively diagonal in the position and momentum basis
and satisfy
\begin{eqnarray}
{\cal V}_j |n\rangle & = & \exp(-i2\pi n^2 (1-j)/2N)|n\rangle\nonumber\\
{\cal T} |k\rangle & = & \exp(-i2\pi  k^2/2N)|k\rangle
\label{kicks}
\end{eqnarray}
where $j$ is an integer. The operator $U_{cat}$ 
can be simply thought of as the 
evolution operator of a kicked system with a Hamiltonian in which 
kinetic and momentum terms are turned on and off alternatively. 
The parameters $a$ and $b$ are integers that measure the 
strength of the potential kicks. It is straightforward to find out what
is the classical system to which the above operator corresponds: 
One way to do this is to take the matrix elements of (\ref{operatorcat}) 
in the computational basis and show that 
\begin{equation}
\langle n'|U_{cat}|n\rangle=K\exp(i2\pi(an^2+bn'^2-2nn')/2N)
\end{equation}
where $K$ is a normalization constant. The exponent in this equation
can be interpreted as the classical action of the system, where 
$n$ and $n'$ the final and initial values of the coordinates. 
Therefore, the classical equations of motion corresponding to 
this system are
\begin{equation}
n= b n' + p'\qquad p= (ab-1)n' + ap'.\label{classicalcat}
\end{equation}
This classical system has been extensively studied \cite{Arnold}: 
For integer values of $a$ and $b$, it is a member of the famous 
family of ``cat'' maps \cite{Arnold}, i.e. all linear 
automorphisms of the torus. The system is chaotic when the 
eigenvalues of the linear transformation ${\cal M}$ mapping 
$\alpha'=(n',p')$
onto $\alpha=(n,p)$ as in (\ref{classicalcat}) are both real (otherwise
the map is integrable). This is the case
when when $Tr{\cal M}=a+b> 2$ (notice that ${\cal M}$ has unit
determinant). In particular, when $a=2$ and $b=1$ this is
the so--called Arnold--cat map ($n=n'+p'$, $p=n'+2p$. This special
cat, according to (\ref{operatorcat}), simply has a kinetic kick 
followed by a potential kick where the potential is an upside-down 
harmonic oscillator. 

The reason why quantum and classical evolution are identical in 
this case can be shown by using our previous results. In fact, 
we need to compute the matrix $Z_{\alpha\beta}$ that evolves
the Wigner function. To do this calculation it is useful to 
first notice that, if the evolution operator $U$ is the one
given in (\ref{operatorcat}), then the following identity holds:
\begin{equation}
U_{cat}\ \hat A(\alpha) = \hat A({\cal M}\alpha)\ U_{cat},
\end{equation}
where, as above, the linear transformation is the one 
given in (\ref{classicalcat}). This means that the unitary 
evolution operator maps the phase space point operator in the same 
way as the classical dynamics does with the phase space point. 
Using this, it is simple to show that the Wigner function 
evolves classically:
\begin{equation}
U=U_{cat} \iff W(\alpha,t+1)=W({\cal M}^{-1}\alpha,t). 
\label{acat}
\end{equation} 

Our result above is not so surprising when seen from the perspective 
of ordinary quantum mechanics of continuous systems. In fact, it is 
well known that quantum and classical evolutions are identical in 
phase space if and only if the classical system has a Hamiltonian 
that is quadratic in $p$ and $q$ (i.e. a harmonic oscillator). 
Here, we showed that the same result is valid for systems with 
a finite dimensional Hilbert space. It is worth pointing out that
it is possible to design a simple quantum circuit to implement the
evolution operator $U_{cat}$ \cite{Shepelyansky}. 
This can be done by noticing that 
the potential kick is diagonal in the computational basis and to 
implement it one simply needs a circuit with the same complexity 
than the Fourier transform (in this way one can construct 
a circuit implementing a controlled phase gate where 
state $|n\rangle$ acquires a phase that depends quadratically 
in $n$ \cite{CircuitsUS}). The operator corresponding to the
kinetic kick con be implemented by a similar circuit in between 
a Fourier transform and its inverse.

\subsection{Boolean gates in phase space}

The family of quantum cat maps is rather large but does not
contain some operators that are more natural from the point
of view of quantum computation. We study here such operators by 
analyzing when does quantum and classical evolution are 
identical in phase space if the dynamics is generated by an 
arbitrary boolean gate. This kind of operation, which is 
a permutation of the computational basis, is an essential
ingredient in Shor's factoring algorithms and others 
\cite{Chuang-Nielsen}.  
So, let us consider a one to one function $f: Z_N\rightarrow Z_N$ 
($f$ is a 
permutation of the first $N$ integers). Given $f$ we can define a 
unitary operator $U_f$ whose action in the computational basis is
to permute vectors in the same way $f$ permutes numbers:
\begin{equation}
U_f|n\rangle=|f(n)\rangle\label{U_f}
\end{equation}

The Boolean function $f$ corresponds to some general (reversible) 
classical circuit. Moreover, one can also associate with $f$ a 
classical map in the phase space. Such map is the one permuting the 
vertical strips corresponding to the different positions according to 
the function $f$, i.e. it maps the vertical strip labeled by 
$n$ into the one labeled by $f(n)$). Here, we want 
to determine under what condition the phase space evolution 
associated with $U_f$ is identical to the classical one (i.e., 
to the mapping of the vertical strips). The answer to this question 
is remarkably simple: the quantum map is identical to the classical 
one if and only if $f(n)=n+a$ (mod$(N)$, for any integer $a$). 
That is to say, 
quantum and classical permutations are identical only in the case of 
$f$ corresponding to a cyclic shift. 

The proof of the above statement is simple: We will consider the 
action of $U_f$ on two classes of states. First we will look at how 
$U_f$ transforms the Wigner function of a computational 
state (i.e., a density matrix like $\rho=|n_0\rangle\langle n_0|$, 
shown in Figure 2 (a)). Then, we will analyze how $U_f$ modifies the 
Wigner function of the interference term between two computational 
states (i.e., a density matrix of the form $\rho=|n_0\rangle\langle n_1|
\pm|n_1\rangle\langle n_0|$ shown in the interference pattern seen in 
Figure 2b). These operators 
form a complete basis of the space of hermitian matrices. Therefore, 
by looking at the evolution of each of these states we are also able 
to understand the evolution of the most general state. As $U_f$ 
simply permutes computational states, it is straightforward to show
that the Wigner function of any such state (shown in Figure 2a) will 
be transformed according to the classical map, i.e. the resulting
Wigner function is obtained from the original one just by permuting
vertical strips according to the classical map $2n\rightarrow f(2n)$). 
However, the fate of a state corresponding to the superposition 
between two computational states is drastically different: As 
shown above, 
the interference fringes are originally located at the midpoint 
between the two interfering states (i.e. at point $n=n_0+n_1$). 
Therefore, for 
the state to evolve classically one needs to impose that this 
midpoint is mapped into the midpoint of the transformed states (i.e. 
that $2f(n_1+n_2)=f(2n_1)+f(2n_2)$). Moreover, the original state has
oscillations with a wavelength that depends on the distance between 
the two computational states ($\lambda=2N/(n_0-n_1)$. Therefore, 
for the state to evolve classically these fringes must remain 
unchanged and the only possibility for this to happen is that 
the distance between the states remains constant (i.e. that 
$2(n_0-n_1)=f(2n_0)-f(2n_1)$). The only solution for these two 
constraints is that $f(n)=n+a$, i.e. a linear function corresponding
to a rigid shift in the computational basis. 

This result can be generalized as follows: Instead of defining $U_f$
as in (\ref{U_f}) we can consider a more general operator $U_{f,g}$ 
which is now associated with two integer one to one functions in such a 
way that $U_{f,g}|n\rangle =\exp(i2\pi g(n)/N) |f(n)\rangle$. Following the 
same steps described above one can show that the operator acts 
in the same way as the corresponding classical map if and only if 
both functions are such that $f(n)=n+a$, $g(n)=n+b$. In such case 
this operator is nothing but the phase space translation $\hat T(a,b)$. 
Thus, our result shows that if one defines the quantum analog of a
classical gate as we did, the only such unitary operator which generates
the same phase space dynamical evolution as its classical analog is 
the one corresponding to a linear rigid shift. 

It is interesting to notice that the structure of the above 
proof (that focuses on how quantum interference terms are 
affected by quantum or classical 
evolution) could also be applied to ordinary quantum mechanics
(of continuous systems) and sheds new insight on the reason why 
only linear systems evolves classically for all times. We 
remark that this result has remained 
unnoticed for quantum computers and may offer some help in our 
efforts to understand the differences between these systems and 
their classical counterparts.

\subsection{Nonlocal (quantum) evolution in phase space}

Apart from the above simple examples with direct classical 
counterpart, most of unitary operators generate a rather
complicated nonlocal evolution for the Wigner function. As an 
example we can explicitly compute matrix elements of 
$Z_{\alpha\beta}$ for some simple case: If we consider a bit flip of
the least significant qubit (which is associated with $\sigma_x^{(0)}$)
we can show that:
\begin{equation}
U=\sigma_x^{(0)} \iff Z_{\alpha 0}=Z_{0\alpha}= 2/N \quad{\rm if}
\quad\alpha\quad {\rm is\quad even}
\end{equation} 
and is equal to zero otherwise ($\alpha=(q,p)$ is even when both
$q$ and $p$ are even). Other matrix elements of $Z_{\alpha\beta}$
are simple to compute but this is enough for the purpose of exhibiting
the nonlocal nature of the evolution. The above equation implies that
when $U=\sigma_x^{(0)}$ the next value of $W(0)$ is proportional to 
the sum of $W(\alpha)$ over all even values of $\alpha$ which is 
indeed a rather nonlocal mapping. 

\begin{figure}
\centering \leavevmode
\epsfxsize 3.2in
\epsfbox{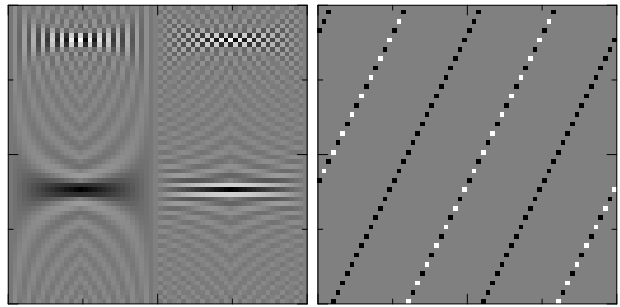}
\vspace{0.25cm}
\caption{Diffraction effects in the Wigner propagation. 
Horizontal (vertical) axis correspond
to position (momentum) basis. In 
the left plot we show the resulting Wigner function obtained
by applying the operator $U_{FT}^{(N/2)}$ (corresponding to a 
Bernoulli shift) to a computational state. The classical 
propagation would contract and rotate the line shown in Fig.2, 
but the quantum propagation shows strong diffraction effects. 
In the right we show the result of propagating the same 
initial state with the `cat' map (equation (\ref{operatorcat})). 
In this case the Wigner function evolves classically and 
shows no diffraction effects.}
\label{halfFT}
\end{figure}

Another simple example of nonlocal evolution is constructed by 
considering a unitary operator $U_{FT}^{(N/2)}$ which consists of 
doing nothing to the most significant qubit and applying the 
Fourier transform to the rest (this is a Fourier transform on a 
space of dimension $N/2$). In such case, the resulting evolution bears
some relation to a $90$ degrees rotation (as the complete Fourier 
transform does) but is a rather nonlocal beast. 
It can be seen that this operation can be understood as the 
quantization of a Bernouilli shift. In fact it is an essential 
ingredient in the construction of the quantum bakers map 
(see \cite{Balasz,Voros,Schack} for more details). Here we are only 
concerned with the phase space description of this operation. 
Classically, when operating on a vertical strip, i.e. a computational
state - this circuit shifts it by $n'\to 2n(mod N)$, contracts 
it to half height
(and double thickness) and rotates by $90^0$. Remnants of this 
behavior can be seen on the left plot of Figure \ref{halfFT} 
where the operation has acted on the computational state 
illustrated in Figure 2. Notice that the phase space representation 
shows strong {\it diffraction } effects which lead to a non 
local $Z_{\alpha \beta}$. It is not difficult to understand 
where these effects come from: the unchanged qubit divides the phase 
space in two regions and the map acts discontinuously on them, thus 
creating the same effect as a screen in an optical system. This is 
contrasted in the right part of Figure \ref{halfFT} with the 
action of Arnold's cat map on the same state. In such case, the 
vertical strip is simply transported by the  classical linear flow.

\subsection{Quantum algorithms in phase space}

\begin{figure} 
\centering \leavevmode 
\epsfxsize 3.2in 
\epsfbox{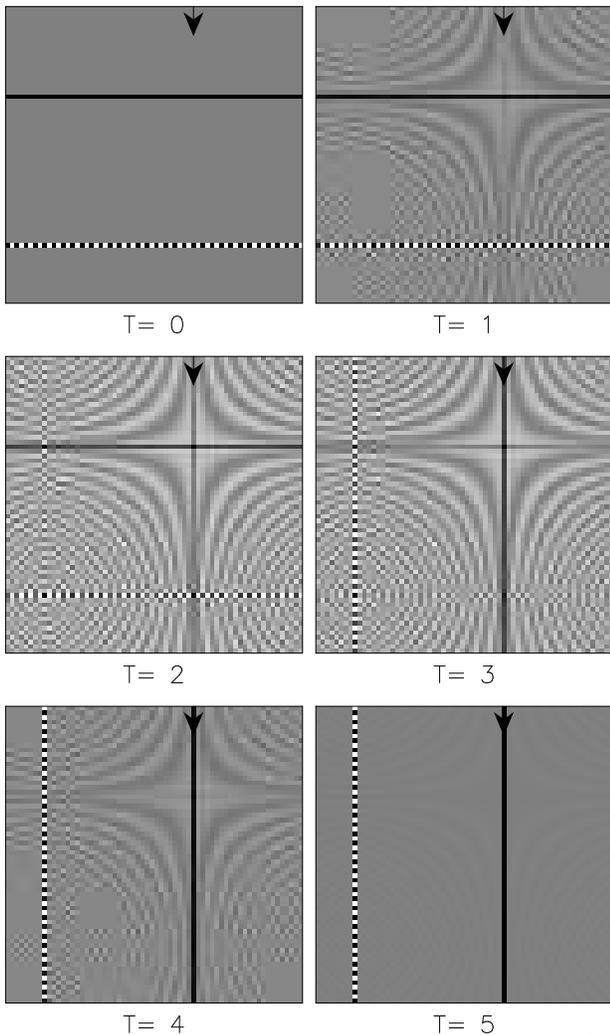} 
\caption{Grover's algorithm in phase space. 
Horizontal (vertical) axis correspond to position (momentum) basis.
After only five iterations 
the initial state (a momentum eigenstate) is transformed into a 
position eigenstate located precisely at the position of the marked 
item (see text).} 
\label{grover}
\end{figure}

Quantum algorithms are nothing but unitary operators that 
have been designed so that after applying them several times, 
they produce a state that encodes the answer to some 
computational problem. 
A measurement on such state gives us information about this answer. 
Quantum algorithms are designed in such a way that
they have a regular behavior as function of $N$. Thus, at least
in some cases, one should be able to 
observe this regularities in a phase space representation. 
In fact, such representation becomes more useful also in the 
large $N$ limit, which is the semiclassical limit (since $1/N$ is 
an effective Planck constant). The study of the phase space
representation of quantum algorithms is only beginning but there
are already some interesting results showing that some 
algorithms may find phase space as a natural arena to be 
represented in. The example we would like to point out here 
is Grover's search algorithm whose phase space representation 
is shown in Figure \ref{grover}

In this Figure we display the Wigner function of the   
quantum state of a computer after every iteration of Grover's search  
algorithm. Our system has an $N=32$--dimensional Hilbert space (i.e.,   
the computer has just $5$ qubits) and the algorithm is designed to   
search for the marked item which we chose here to be $q=16$ (indicated with   
an arrow in the plot). The initial state is chosen to be an equally   
weighted superposition of all computational states which, for the  
purpose of making the plots more visible, we chose to be a non-zero  
momentum state (the usual choice for the initial state  
in Grover's algorithm is a zero  
momentum state but the algorithm works as well with an initial state  
$|k\rangle$ with nonzero momentum). 
The quantum search algorithm consists of the iteration of 
a unitary operator $U_G$ which can be decomposed as $U_G=U_k U_o$. 
The operator $U_o$ represents a 
call to an oracle that only distinguishes the marked 
item $\omega$ from the rest: $U_o=I-2|\omega\rangle\langle\omega|$. The
operator $U_k$ (the inversion about the mean) has no information 
about the marked item and is chosen as $U_k=I-2|k\rangle\langle k|$. 
The quantum computer should be initially prepared in the state 
$|k\rangle$. After a number of iterations $T\approx\pi\sqrt N/4$ 
the probability of detecting the position $q=\omega$ becomes of 
order unity (with small correction decaying as $1/N$). Thus, 
by measuring the position at this time one would discover 
the marked item. From the figure one clearly observes that this 
algorithm is very  
simple when seen from a phase space representation. The initial state   
has a Wigner function which is a horizontal strip (with its   
oscillatory companion). After each iteration $W(q,p)$ shows a
very simple Fourier-like pattern and becomes a pure coordinate state at the
end of the search (in our case, the optimal number of iterations 
is $T\approx\pi\sqrt N/4\approx 5$).
This representation shows that, as a map, Grover's algorithm
has a fixed phase space point with coordinate equal to the marked item ($q=16$
in our case) and momentum equal to the one of the initial state.

It is interesting to speculate if this phase space representation
may be of guidance to create new quantum algorithms or improve on 
existing ones. We cannot present clear evidence to support this but
we believe this will be the case. However, some careful thought is
required. At this moment we can only offer the reader some of our
experience on the way one should not proceed: 
Viewing Figure \ref{grover}
one may be tempted to try to find a quantum map that has the same
fixed point as the above but transform horizontal strips into vertical
ones faster than Grover's. It is interesting to notice that such 
maps do exist. In fact, one just needs a hyperbolic (instead of 
an integrable) map with the same fixed point structure. 
In fact, a hyperbolic map with the same fixed point structure 
would perform the same operation {\sl exponentially} faster, 
because it would transform a state initially along the stable direction 
into one along the unstable one in a time of order $\log(N)$ . Such 
hyperbolic maps can be easily constructed as variations of the quantum 
bakers map. However, their construction requires from the oracle much 
more information than a simple yes/no query and thus do
not qualify as fast search algorithms.

\section{Measuring the Wigner Function}

In this section we review a general procedure to directly 
measure the value of the Wigner function of a system in any 
phase space point (the procedure applies for discrete 
and also for continuous systems). This measurement strategy 
was originally proposed in \cite{NatureUS} as an application of 
a very general tomographic scheme. It generalizes previous proposals
that were put forward to measure the Wigner function in 
quantum optics or cavity QED setups (see \cite{Davidovich00}
for a review). 

\begin{figure}
\epsfxsize=7.6cm
\epsfbox{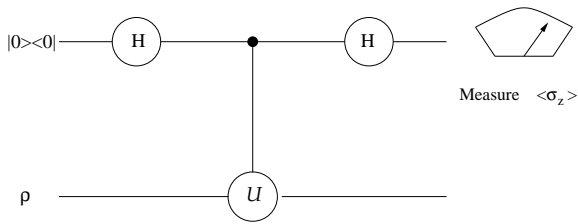}
\vspace {0.5cm}
\caption{Circuit for measuring $Re(Tr( U\rho))$, for a given 
unitary operator $U$.}
\label{wcircuit}
\end{figure}

The basic idea of the measurement is best described in terms 
of the quantum algorithm represented by the simple quantum circuit 
shown in Figure \ref{wcircuit}. It works 
as follows: Suppose you want to measure
the Wigner function of a system that is prepared in some quantum 
state $\hat\rho$. What you need to do is to bring this system in contact 
with an ancillary qbit prepared in the state $|0\rangle$ and then
``run'' the quantum computation described by the circuit shown in 
Figure \ref{wcircuit}. The  ancillary qbit plays the role of 
a ``probe particle'' in a 
scattering--like experiment. The algorithm represented in the 
circuit is: i) Apply an Hadamard transform to the ancillary qbit 
(where $H|0\rangle=(|0\rangle+|1\rangle)/\sqrt 2$, 
$H|1\rangle=(|0\rangle-|1\rangle)/\sqrt 2$), ii) Apply 
a ``controlled--$\hat U$'' operator (if the ancilla is in state
$|0\rangle$ this operator acts as the identity for the system but if 
the state of the ancilla is $|1\rangle$ it acts as the 
unitary operator $\hat U$ on the system), iii) Apply another 
Hadamard gate to the ancilla and finally perform a {\it weak} 
measurement on this qbit detecting its polarization (i.e., measuring 
the expectation values of Pauli operators 
$\sigma_z$ and $\sigma_x$). It is easy to show that the above 
circuit has the following remarkable property: 
\begin{equation}
\langle\sigma_z\rangle=Re(Tr(\hat U\hat\rho)), \quad
\langle\sigma_y\rangle=-Im(Tr(\hat U\hat\rho)).
\end{equation}
Thus, the final polarization measurement reveals a property 
determined both by the initial state $\hat\rho$ and the 
unitary operator $\hat U$.

It is clear that this circuit can be used for dual purposes: On 
the one hand, we can use it to extract information on the operator 
$\hat U$ if we know the state $\hat\rho$. On the other hand, 
the same circuit can be used to learn about the state $\hat\rho$ by 
using some specific operators for $\hat U$. In this sense, this circuit can
be adapted to act either as a tomographer or as an spectrometer which,
therefore, are two dual faces of the simple quantum computation  
shown above \cite{NatureUS}. The analogy with a typical scattering 
experiment is quite clear: the probe particle is prepared in a well 
determined state, then it interacts with the system. From the measurement
of some statistical distribution of the probe (in this case the polarization)
we are able to infer properties of either the interaction or the state
of the scatterer. 

In \cite{NatureUS} we showed that this circuits can be used to 
measure the Wigner function. This is done by simply choosing 
$\hat U= \hat A(\alpha)$. By doing this, every time one runs the
algorithm a direct measurement of $W(\alpha)$ is obtained (for any given 
phase space point $\alpha$). This measurement strategy
can be, in principle, used both for the continuous and the discrete case. 
For example, in the continuous case, to measure $W(\alpha)$ of 
a quantum particle one has
to run the algorithm with the system in the initial state 
$D(\alpha)\hat\rho D^\dagger(\alpha)$ (obtained by displacing $\hat\rho$) 
and using $\hat U=\hat R$ (it is clear that this is equivalent, but
simpler, than applying the circuit directly with 
$\hat U=\hat A(\alpha)$). 
The measurement of the Wigner function for the discrete case can be 
done in a similar way but in that case it is more convenient 
to directly use 
$\hat U=\hat A(\alpha)$. In both cases, as the operator $\hat A(\alpha)$ 
is both unitary and hermitian, the measurement of the Wigner function 
only requires determining the $z$--component of the polarization. 

Measuring directly the Wigner function of a quantum system has been 
the goal of a series of experiments proposed and performed 
in various areas of physics (all dealing with continuous systems, 
\cite{WignerMeas}). It is interesting to notice that the above 
measurement strategy precisely describes the experiment recently performed 
to determine the Wigner function of the electromagnetic field in a 
cavity QED setup \cite{DavidovichLut,Harocheetal}. In fact, in such case 
the system is the mode of the field stored in a high--$Q$ 
cavity and the ancillary qbit is a two level atom. Remarkably, 
this experiment can be easily described in terms of the 
application of the algorithm shown in Figure \ref{wcircuit}: 
i) The atom goes through a Ramsey zone that has the effect of 
implementing an Hadammard transform. A radio-frequency source is 
connected to the cavity displacing the field in phase space (by an 
amount parametrized by $\alpha$),
ii) The atom goes through the cavity interacting dispersively with 
the field. The interaction is tuned in such a way that, if the atom 
is in state $|g\rangle$ nothing happens but if the state
of the atom is $|e\rangle$ the state acquires a phase shift 
of $\pi$ per photon in the cavity. So, this interaction 
is simply a controlled--$\exp(-i\pi\hat N)$ gate (where $\hat N$ 
is the photon number operator) which is nothing but a controlled 
reflection. iii) The atom leaves the cavity entering a new Ramsey zone 
that enforces a new H--gate. Finally the atom is detected in a counter
either in the ground $|g\rangle$ or in the excited $|e\rangle$ state. 
The experiment is repeated many times and the 
Wigner function is obtained as the difference between the probability 
of the atom being in the excited and ground 
states: $W(q,p)= 2 (P(e)-P(g))/\hbar$. As we see, this cavity--QED 
experiment is an example of a tomographic experiment represented
by the circuit of Figure \ref{wcircuit}. 
The important point is that the method
is generalizable to arbitrary systems (continuous or discrete) 
when viewed in terms of the above circuit. 
  
For the discrete case, one can show that the circuit in 
Figure \ref{wcircuit} can be efficiently decomposed as a 
sequence of elementary gates. 
For this, one only needs to implement controlled-$\tilde U$, 
$\tilde V$ and $\hat R$ operations. 
All these can be implemented via efficient networks like the ones shown 
in \cite{networks} which require a number of elementary gates that depends
polinomially on $log(N)$. For example, to decompose $\hat V$ 
as a sequence of one qbit gates we can notice that its action on 
a computational state $|n\rangle$ is:
\begin{equation}
\hat V^p |n\rangle = \prod_{i=0}^{L-1} \left[ \exp( i 2 \pi  p 2^i/N ) 
\right]^{n_i} |n\rangle \nonumber
\end{equation}
where $n_i$ is the binary expansion of the number $n$. From this 
expression it is clear that to implement $\hat V$ we just have 
to act independently on each qbit. The operator $\phi_i$ 
acting on the i--th qbit should be such that 
$\phi_i|0\rangle=|0\rangle$ and 
$\phi_i|1\rangle=\exp(i2\pi p 2^i/N)|1\rangle$.
On the other hand, to implement $\hat U$, which is a shift in 
the position basis, we can either use a circuit which 
performs addition mod $N$ as described in \cite{networks} or
notice that a circuit for $\hat U$ is obtained from the
one corresponding to $\hat V$ by using 
$\hat U= U_{FT}^{-1}\hat V U_{FT}$. 
Finally, to implement the reflection operator we use the identity
(\ref{R=UFT^2}) which states that two 
Fourier transforms are equivalent to a reflection. 
Therefore, the complete network can be obtained straightforwardly. 

\begin{figure}[h]
\epsfxsize=8.6cm
\epsfbox{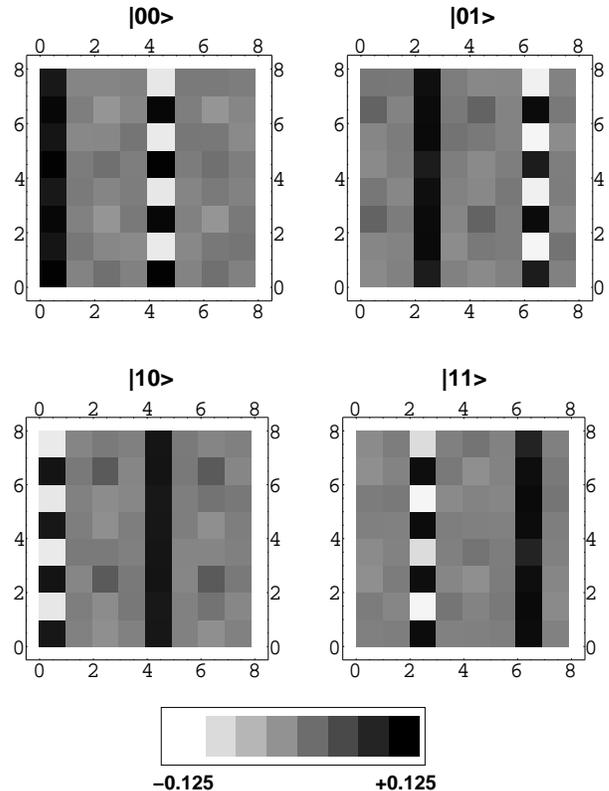}
\caption{Measured Wigner functions for the four computational states 
of a two--qbit system (built with a liquid sample of TCE 
in an NMR spectrometer). Horizontal (vertical) axis correspond
to position (momentum) basis. Ideally, these 
Wigner functions should be nonzero only on two vertical strips where they
take values which are $\pm 1/8$. Experimental results show small deviations
from these values (with a maximum error of $15\%$.}
\label{wmeasurement}
\end{figure}

For small values of $N$ the resulting networks  
are indeed very simple. In \cite{NatureUS} we have presented 
the results of the measurement of the 
Wigner function for the particular case of $N=4$ (two qbits) for a variety
of initial states. In this case the reflection operator is simply 
a controlled
not (CNOT) gate (where the control is in the least significant qbit). 
The circuit for $\hat U$ is the same CNOT followed by bit flip in the 
control. Analogously, the circuit for $\hat V$ is a sequence 
of controlled phase gates. Thus, the complete measurement circuit 
has at most one Tofolli gate and several two qbit gates. We 
experimentally measured the Wigner function of the four computational
states of a two qbit quantum computer. 
The ideal result is shown in Figure 2 and the measured 
Wigner functions are seen in \ref{wmeasurement}. 

To perform these experiments 
we used a liquid sample of trichloroethylene (TCE) 
dissolved in chloroform. This molecule has been used in several 
three qubit 
experiments where the proton ($H$) and two strongly coupled  $^{13}C$
nuclei ($C_1$ and $C_2$) store the three qbits \cite{nmrexperiments}. 
In our case, we used $C_1$ as our control qbit (the probe particle 
in Figure \ref{wcircuit}) and the 
pair $H$-$C_2$ to store the state whose Wigner function we tomographically
measure. The measured coupling constants 
are $J_{HC_1}=200.76 \ \mbox{Hz}$, $J_{HC_2}=9.12 \ \mbox{Hz}$ 
and $J_{C_1C_2}=103.06 \ \mbox{Hz}$ while the $C_1$-$C_2$ chemical 
shift is $\delta_{C_1C_2}=908.88 \ \mbox{Hz}$. To evaluate the 
Wigner function in each of the independent $16$ phase space points
we expressed the algorithm of Figure \ref{wcircuit} as a sequence
of r.f. pulses and delays (the number of pulses in each sequence 
depends on the phase space points and varies between $5$--$20$ and 
take at most $100$ms to execute ($\ll T_1,T_2$ of our sample)).  
Experiments were done at room temperature and we used temporal averaging
to distill the initial pseudo-pure states \cite{temporalav}. 
The experiments were done on a standard $500$Mhz NMR spectrometer
(Bruker AM-500 at LANAIS in Buenos Aires and DRX-500 at Los Alamos). 
We used a $5$mm probe tuned to $^{13}C$ and $^1H$ frequencies 
of $125.77$ MHz and $500.13$ Mhz, respectively. The measured Wigner 
functions shown in figure \ref{wmeasurement} agree with the ideal 
one within (in the worst case) approximately 
$15\%$ error. The most important sources of errors are understood and 
come from the effects of strong
coupling (that alter the behavior of our gates) and 
numerical uncertainty in 
integrating the spectra. It is clear that the result of this 
simple experiment, 
that illustrate the tomographic measurement of a discrete Wigner function, 
agrees very well with the theoretical expectation. 

\section{Conclusion}

In this paper we first reviewed the formalism to represent in 
phase space the states and the evolution of a quantum system 
with a finite dimensional Hilbert space. Then, we applied this
formalism to study the phase space representation of quantum 
computers and quantum algorithms. Finally, we discussed how to 
perform direct measurements to determine the Wigner function. 
Our approach, based on the use of phase space point operators 
to define the Wigner function enables us to find simple 
proofs for a variety of old and new results concerning both 
states and algorithms.

The differences between quantum and classical evolution in phase
space are clearly seen using this phase space representation: For
the Wigner function to evolve classically it is necessary (and 
sufficient) that the phase space point operators are transformed
by the unitary evolution operator in the same way that classical 
evolution maps phase space points. In our paper we described a
rather interesting family of operators with this property. It 
contains all linear homeomorphisms in the torus (i.e., the family
of all ``cat'' maps) which can be efficiently 
represented by simple quantum circuits (since they correspond to 
kicked systems where kinetic and potential terms act alternatively). 
Moreover, this family also includes
all cyclic shifts in the computational (or the momentum) basis, 
reflections and the Fourier transform. 
At a deeper level all these properties are a consequence of the well 
known invariance of the Wigner function under the action
of the unitary representation of linear canonical transformations
(that form the metaplectic group). Other unitary maps, that 
have a definite classical non--linear counterpart, will display
for some time an almost classical behavior, signalled by a sharply peaked 
$Z_{\alpha \beta}$ matrix. This is the semiclassical limit, that will 
be reached in the limit $N\to\infty$, which is an 
important regime for interesting 
quantum algorithms since it corresponds to the limit of large 
number of qubits. A much wider category of maps can display a more drastic 
wave phenomenon: strong diffraction due to discontinuities. Consider, as 
the simplest example the phase space operation of a control-$U$ gate. 
The controlling qubit divides phase space in two disjoint regions and 
the $U$-gate acting on the remaining qubits affects these two regions 
differently. In this sense it acts as a prism or a screen 
in an optical system. 
Another simple example of this sort is the action of the operator 
$U_{FT}^{N/2}$, that does not act on the most significant
qbit but applies the Fourier transform to the rest (the diffraction 
effects are clearly seen in Figure \ref{halfFT}). The implications 
and usefulness of these semiclassical features in quantum algorithms
remain largely unexplored and will be the subject of further investigation
\cite{algorithmsUS}.
 

\acknowledgments
We thank useful discussions with R Laflamme, E. Knill, C. 
Negrevergne. We also benefited from our interaction with 
P. Bianucci. JPP thanks L. Davidovich for his hospitality in Rio 
and his insight. This work was partially supported with funds from 
Ubacyt, Conicet, Anpcyt and Fundaci\'on Antorchas.


\begin{references}

\bibitem{Wigner} M. Hillery, R.F. O'Connell, M.O. Scully, E. P. 
Wigner, {\it Phys. Rep.}{\bf 106} (1984), 121.

\bibitem{WignerUS} P. Bianucci, C. Miquel, J. P. Paz and M. Saraceno,
{\it Discrete Wigner functions and the phase space representation of
a quantum computer} quant-ph/0105091, submitted to PRL (2000).

\bibitem{Davidovich00} L. Davidovich {\it Quantum optics in cavities, 
phase space representations and the classical limit of quantum 
mechanics}, in {\it New perspectives on quantum mechanics. AIP Conf. Proc.} 
{\bf 464}, ed S. Hacyan et al. (1999, New York, AIP); also in 
Proccedings of the First PASI Conference on Chaos, 
Decoherence and Entanglement; (2000) available at http://kaiken.df.uba.ar. 

\bibitem{decowigner} see for example W. H. Zurek, {\it Physics Today}  
{\bf 44}, (1991) N 10, 36;  
(1994) 2508;   
J. P. Paz, S. Habib and W. H. Zurek, {\it Phys. Rev.} {\bf D47} 
(1992) 488.   

\bibitem{PazZurek00} see J. P. Paz and W. H. Zurek, {\it 
``Environment induced superselection and the transition from 
quantum to classical}; (2001). in {\it ``Coherent 
matter waves, Les Houches Session LXXII};, edited by R Kaiser, 
C Westbrook and F David, EDP Sciences, Springer Verlag (Berlin) (2001)
533-614 and references therein. 

\bibitem{Chuang-Nielsen} {\it "Quantum Information and Computation"}, 
I. Chuang and M. Nielsen (2000), Cambridge University Press.

\bibitem{Wooters} W. K. Wooters, Ann. Phys. NY {\bf 176} (1987), 1

\bibitem{Leonhardt} U.Leonhardt, Phys. Rev. Lett. {\bf 74}, 4101 (1995). 
U.Leonhardt, Pys. Rev. A {\bf 53}, 2998 (1996).

\bibitem{Hannay} J. H. Hannay, M. V. Berry, Physica {\bf 1D}, 267 (1980)

\bibitem{Rivas} A. Rivas, A. M. Ozorio de Almeida, Ann.Phys. {\bf 276}
(1999), 123

\bibitem{Bouzouina} A. Bouzouina, S. De Bievre, Comm. Math. Phys. {\bf 178}
(1996)83

\bibitem{Bertrand} J.Bertrand and P.Bertrand, Found. Phys. {\bf 17} (1987) 
397.

\bibitem{Fano}U.Fano, Review of Modern Physics, {\bf 29} (1957) 75.

\bibitem{Schwinger} J. Schwinger, {\it Proc. Nat. Acad, Sci.}{\bf 46} (1960),
570, 893. 

\bibitem{CircuitsUS} J. P. Paz and M. Saraceno, (2001) in progress.

\bibitem{Arnold} V.I. Arnold and A. Avez, {\it Ergodic 
Problems in Classical Mechanics}, Benjamin, New York, (1968).

\bibitem{Shepelyansky} B. Georgeot and D. L. Shepelyansky, {\it 
Phys. Rev. Lett.} {\bf 86} (2001) 2890-2893; see also B. Georgeot and D. L. 
Shepelyansky, {\it Phys. Rev. Lett.} {\bf 86} (2001) 5393-5396.

\bibitem{Balasz} N.L. Balasz and A. Voros, {\it Ann. Phys.} 
{\bf 190} (1990) 1.

\bibitem{Voros} A. Voros y M. Saraceno, 
{\it Physica} {\bf D79} (1994) 206; see also M. Saraceno, {\it Ann. Phys.} 
{\bf 199} (1990) 37.

\bibitem{Schack} R. Schack, {\it Phys. Rev} {\bf A57} (1998) 1634-1635;
R. Schack and T. Brun, {\it Phys. Rev.} {\bf A 59} (1999) 2649-2658; 
see also R. Schack and Caves C. M.,
{\it Appl. Algebr. Eng. Comm.} {\bf 10} (2000) 305-310; 
A. N. Soklakov and R. Schack, {\it Phys. Rev.} {\bf E 61} (2000)
5108-5114. 

\bibitem{NatureUS} C. Miquel, J.P. Paz, M. Saraceno, E. Knill, 
R. Laflamme, C. Negrevergne (2001) submitted to Nature. 

\bibitem{WignerMeas} T. J. Dunn {\it et al.}, Phys. Rev. Lett. 
{\bf 74} (1994)
884; D. Leibfried {\it et al.} Phys. Rev. Lett. {\bf 77} (1996) 4281;
see also Physics Today {\bf 51} no. 4 (1998) 22.

\bibitem{DavidovichLut} L. G. Lutterbach and L.
Davidovich, Phys. Rev. Lett. {\bf 78} (1997) 2547;
Optics Express {\bf 3} (1998) 147.

\bibitem{Harocheetal} G. Nogues et al, Phys. Rev. {\bf A} (2000); see
also X. Maitre et al, Phys. Rev. Lett. {\bf 79} (1997) 769.

\bibitem{networks} C.Miquel, J.P.Paz, R.Perazzo, Phys. Rev. A {\bf 54}
(1996) 2605. 

\bibitem{nmrexperiments} see for example:
M. A. Nielsen, E. Knill, R. Laflamme,
Nature {\bf 396} (1998), 52--55; D. G. Cory {\it et al.},
Phys. Rev. Lett. {\bf 81} (1998) 2152

\bibitem{temporalav} E.Knill, I.Chuang and R.Laflamme, 
Phys. Rev. A {\bf 57} 3348.

\bibitem{algorithmsUS} J. P. Paz and M. Saraceno, (2001) work 
in progress. 

\end{references}
\end{document}